\documentclass[12pt,a4paper,notitlepage]{article}
\usepackage{amsmath,amssymb,amsthm}
\usepackage{citesort} 

\newcommand{\shft}[2]{#1^{{}^{(#2)}}}
\newcommand{\cas}[1]{\mathop{\mathrm{cas}}\left(#1\right)}
\newcommand{\wcas}[1]{\widetilde{\mathrm{cas}}\left(#1\right)}
\newcommand{\bcas}[1]{\overline{\mathrm{cas}}\left(#1\right)}
\newcommand{\hdcas}{\mathrm{cas}}
\newcommand{\hdwcas}{\widetilde{\mathrm{cas}}}
\newcommand{\diag}{\mathrm{diag}}
\newtheorem{theorem}{Theorem}

\newtheorem{remark}{Remark}

\newtheorem{example}{Example}

\title{\bf On the Toda Lattice Equation \\
  with Self-Consistent Sources }

\author{Xiaojun Liu\thanks{E-mail:lxj98@mails.tsinghua.edu.cn} \quad
  and \quad Yunbo Zeng\thanks{E-mail:yzeng@math.tsinghua.edu.cn}\\
  Department of Mathematical Sciences,\\ 
  Tsinghua University, 100084,\\ 
  Beijing, PRC.}  

\date{} 

\begin{document}

\maketitle

\begin{abstract}
  The \emph{Toda lattice hierarchy with self-consistent sources} and
  their Lax representation are derived. We construct a \emph{forward}
  Darboux transformation (FDT) with arbitrary functions of time and a
  \emph{generalized} forward Darboux transformation (GFDT) for Toda
  lattice with self-consistent sources (TLSCS), which can serve as a
  non-auto-B\"acklund transformation between TLSCS with different
  degrees of sources. With the help of such DT, we can construct many
  type of solutions to TLSCS, such as rational solution, solitons,
  positons, negetons, and soliton-positons, soliton-negatons,
  positon-negatons etc., and study properties and interactions of
  these solutions.
\end{abstract}

\hskip\parindent

\pagebreak


\section[Introduction]{Introduction}

Recently, soliton equations with self-consistent sources (SESCSs) in
several 1+1 and 2+1 dimensional \emph{continuous} cases, which have
very important applications in many fields, such as hydrodynamics,
solid state physics, plasma physics, etc. have been widely studied
\cite{art:Mel87,art:Mel88-2,art:Mel91,art:LeonLatifi90,art:ShcDok96}.
For example the KdV equation with self-consistent sources describes
the interaction of long and short capillary-gravity waves
\cite{art:LeonLatifi90}. Various methods have been used to construct
their solutions, such as Inverse scattering methods
\cite{art:Mel88-2,art:Mel91,art:ZengMaLin00,art:LinZengMa01}, Darboux
transformation methods
\cite{art:ZengMaShao01,art:ZengShaoMa02,art:ZengShaoXue03,art:XiaoZeng04},
Hirota bilinear methods
\cite{art:Matsu91,art:ZhangDaChenDen03,art:DengChenZhang03} etc.
Comparing with Darboux transformations (DT) for soliton equations,
generalized binary Darboux transformations (GBDT) with arbitrary
functions of time developed in
\cite{art:ZengMaShao01,art:ZengShaoMa02,art:ZengShaoXue03,art:XiaoZeng04}
provides a non-auto-B\"acklund transformations between two soliton
equations with different degrees of sources, and enable us to find
various solutions, such as soliton, positon, negaton, soliton-positon,
soliton-negaton, positon-negaton and etc.

However, in contrast with the continuous case, the integrable discrete
systems with self-consistent sources and their physical applications
have not been studied yet. We will present a way to construct
integrable discrete systems with self-consistent sources basing on the
constrained flows of the integrable discrete system \cite{art:Zeng97}
by regarding the latter as the stationary equation of the former. Then
we derive a \emph{forward} Darboux transformation (FDT) with an
arbitrary function of time for the integrable discrete system with
self-consistent sources by improving the method in
\cite{art:ZengMaShao01,art:ZengShaoMa02,art:ZengShaoXue03,art:XiaoZeng04}.
We use a discrete system, the Toda lattice equation with
self-consistent sources (TLSCS) to illustrate our method. The FDT with
an arbitrary function of time is generated by using a linear
combination of two independent eigenfunctions of the Lax pair with the
combination coefficient explicitly depends on time. It serves as a
non-auto-B\"{a}cklund transformation between two TLSCS with different
degrees of sources. We give a formula for multi-time repeating of such
DT. We also construct the \emph{generalized} FDT (GFDT) with arbitrary
functions of time. The formula for GFDT is quite similar to the
generalized \emph{Wronski} determinant formulae \cite{art:Matveev2002}
and the generalized \emph{Casorati} determinant formulae
\cite{art:StahlhofenMatveev95,art:MarunoMawx2004}.

It is well known that Toda lattice equation possesses rich families of
solutions including rational solution, soliton, positon, negatons and
soliton-positon, soliton-negaton, positon-negaton. Soliton is a fast
decaying pulse like solution without singularity \cite{bk:TodaLatt81}.
It describes a wave propagating in a constant speed. Positon is an
oscillating and slowly decaying solution with singularities. It leads
to a trivial scattering matrix (called "supertransparent") when
inserted as potentials in the finite-difference Schr\"{o}dinger
equation of the corresponding Lax pair \cite{art:StahlhofenMatveev95}.
The interaction between positon and other types of solutions is very
interesting. There is no phase shift for others during the course of
collision. Negatons is another type of solutions with singularities
oscillating but fast decaying and with non-trivial scattering matrix.
Unlike the positon case, there is phase shift for other types of
solutions during the interaction between negaton and others.

In the second part of our paper, we show that TLSCS also have such
types of solutions. These solutions can be obtained easily through
GFDT. Some of these solutions as well as their analytic properties are
presented in our paper. Differences between solutions for TLSCS and
solutions for Toda lattice equation are also studied. In particular, a
new feature regarding negaton-positon (negaton-soliton) interaction is
analyzed.

The paper will be organized as follows. We first review the Toda
lattice hierarchy in section \ref{sec:TLH}. In section
\ref{sec:TLHSCS}, through the nth-constrained flow for Toda lattice
hierarchy we give the Toda lattice hierarchy with self-consistent
sources and its Lax representation. Based on the Darboux
transformation for Toda lattice equation, we develop a method to
construct the FDT and GFDT with arbitrary functions of time for TLSCS
in section \ref{sec:DTTLSCS}. In section \ref{sec:solut-toda-latt}, we
construct some solutions for TLSCS by using GFDT with arbitrary
functions of time. Some properties of these solutions are analyzed
therein.

\section{Toda Lattice Hierarchy}
\label{sec:TLH}
We first review the Toda lattice hierarchy.Assume $f=f(n,t)$ for
$n\in\mathbb{Z}$ and $t\in \mathbb{R}$. Define shift operator $E$ and
difference operator $\Delta$ as follows
\begin{eqnarray*}
  (Ef)(n,t)&=& f(n+1,t)\,,\\
  (\Delta f)(n,t) &=& (E - 1)f(n,t) = f(n+1,t) - f(n,t)\,.
\end{eqnarray*}
We often denote $E^kf(n,t)$ by $\shft{f}{k}$. The inverse of $E$ and
$\Delta$ are defined as
\begin{displaymath}
  (E^{-1}f)(n,t)=f(n-1,t),
\end{displaymath}
\begin{displaymath}
  (\Delta^{-1}f)(n,t)=\left\{
    \begin{array}{cl}
      \sum_{i=0}^{n-1}f(i,t) & n\geq 1;\\
      0& n=0;\\
      -\sum_{i=n}^{-1}f(i,t) & n\leq -1.
    \end{array}\right.
\end{displaymath}

Consider the following discrete isospectral problem
\cite{bk:MatveevSalle91,art:StahlhofenMatveev95}
\begin{equation}
  \label{eq:TodaIsospectralProb}
  L\psi=\shft{(v\psi)}{1}+p\psi+\shft{\psi}{-1}=\lambda\psi,
\end{equation}
where $v=v(n,t)$, $p=p(n,t)$, $\psi=\psi(n,\lambda,t)$. This equation
can be rewritten as the following $2\times 2$ matrix eigenvalue
problem
\begin{equation}
  \label{eq:TodaMatEgnProb}
  \shft{\Psi}{-1} = U(v,p,\lambda)\Psi\,,
\end{equation}
where
\begin{displaymath}
U(v,p,\lambda):= \left(
  \begin{array}{cc}
    0 & 1 \\ 
    -\shft{v}{1} & \lambda-p
  \end{array}
\right)\quad,\quad \Psi:= \left(
  \begin{array}{l}
    \shft{\psi}{1}\\
    \psi
  \end{array}
\right).
\end{displaymath}
First consider the following \emph{stationary zero-curvature equation}
for generating function $\Gamma$ \cite{art:Tu90},
\begin{equation}
  \label{eq:AdjRepr}
  \shft{\Gamma}{-1}U-U\Gamma = 0\,,
\end{equation}
where
\begin{displaymath}
\Gamma:=\sum_{i=0}^{+\infty}\Gamma_i\lambda^{-i}=\sum_{i=0}^{+\infty}
\left(
  \begin{array}{cc}
    a_i & b_i \\
    c_i & -a_i
  \end{array}
\right)\lambda^{-i}\,,
\end{displaymath}
and $a_i$,$b_i$,$c_i$ are functions of $n$ and $t$. The first few of
these read
\begin{alignat*}{3}
  a_0 &= \frac{1}{2},&  \qquad b_0&=0,            & \quad c_0&= 0,\\
  a_1 &= 0,          &  \qquad b_1&=-1,           & \quad c_1&=\shft{v}{1},\\
  a_2 &=\shft{v}{1}, &  \qquad b_2&=-\shft{p}{1}, & \quad c_2&=p\shft{v}{1},\\
  a_3 &=(p+\shft{p}{1})\shft{v}{1},& \qquad
  b_3&=-({\shft{p}{1}}^2+\shft{v}{1}+\shft{v}{2}), &\quad c_3
  &=\shft{v}{1}(p^2+v+\shft{v}{1}).
\end{alignat*}
In general
\begin{align*}
  \Delta a_{i+1} &=\shft{p}{1}\Delta a_i-c_i-\shft{b_i}{1}\shft{v}{2},\\
  b_{i+1} &=\shft{p}{1}b_i-\Delta a_i,\quad c_i =-\shft{v}{1}\shft{b_i}{-1}.
\end{align*}
Define the modification matrix
\begin{displaymath}
\Delta_n:=\diag (b_{n+1}+\delta,\,\delta),
\end{displaymath}
where $\delta$ is an arbitrary constant. Let
\begin{displaymath}
  V_n:=\left(\lambda^n\Gamma\right)_+ + \Delta_n
  =\sum_{i=0}^n\Gamma_i\lambda^{n-i} + \Delta_n,
\end{displaymath}
and 
\begin{equation}
  \label{eq:TLHtFlow}
  -\Psi_{t_n}=V_n\Psi, \quad (n\geq 1).
\end{equation}
Then the compatibility condition of \eqref{eq:TodaMatEgnProb} and
\eqref{eq:TLHtFlow} gives rise to the following zero-curvature
representation of the Toda lattice hierarchy
\begin{equation}
  \label{eq:ZeroCurvRepr}
  U_{t_n}=UV_n-\shft{V_n}{-1}U ,\quad (n\geq 1).
\end{equation}

Hamiltonian form for Toda lattice hierarchy is given by
\begin{displaymath}
  \left(
    \begin{array}{c}
      v\\ p
    \end{array}
  \right)_{t_n}
  =J
  \left(
    \begin{array}{c}
      \frac{\delta H_n}{\delta v}\\ \frac{\delta H_n}{\delta p} 
    \end{array}
  \right)
  =J \left(
    \begin{array}{c}
      \shft{a_{n+1}}{-1}/v\\ -\shft{b_{n+1}}{-1} 
    \end{array}
  \right)
\end{displaymath}
with the density $H_n:= - b_{n+2}/(n+1)$ and Hamilton operator
\begin{displaymath}
J := \left(
  \begin{array}{cc}
    0 & v\left(E^{-1}-1\right) \\
    (1-E)v & 0 
  \end{array}
\right).
\end{displaymath}
For $n=1$ in \eqref{eq:ZeroCurvRepr} we have the well-known Toda
lattice equation
\begin{equation}
  \label{eq:TodaByVP}
  \begin{array}{ccl}
    v_t &=& v \shft{p}{-1} - v p, \\ 
    p_t &=& v - \shft{v}{1}.\\
  \end{array}
\end{equation}
Set $v := \exp \left(\shft{x}{-1}-x \right)$, $p:= x_t$, equations
\eqref{eq:TodaByVP} can be represented as
\begin{equation}
  \label{eq:TodaByODE}
  x_{tt} = \exp \left( \shft{x}{-1}-x \right)-\exp \left( x-\shft{x}{1}\right).
\end{equation}

\section[Toda lattice hierarchy with self-consistent sources]{The Toda
  lattice hierarchy with self-consistent sources}
\label{sec:TLHSCS}

First we briefly review the  \emph{$n$th-constrained flow} for Toda
lattice hierarchy \cite{art:Zeng97} which is defined as the following
system
\begin{subequations}
  \begin{align}
    &\left(
      \begin{array}{c}
        \frac{\delta H_n}{\delta v}\\
        \frac{\delta H_n}{\delta p}
      \end{array}
    \right) + \sum_{j=1}^N \left(
      \begin{array}{c}
        \frac{\delta \lambda_j}{\delta v}\\
        \frac{\delta \lambda_j}{\delta p}
      \end{array}
    \right) =0 \label{eq:ConstraintFlow}\\
    & L \phi_{j+} := \shft{(v\phi_{j+})}{1} + p\phi_{j+} +
    \shft{\phi_{j+}}{-1} = \lambda_j\phi_{j+},
    \label{eq:constrFlowSourceEigenProblem}\\
    & L^* \phi_{j-} := v\shft{\phi_{j-}}{-1} + p\phi_{j-}
    +\shft{\phi_{j-}}{1} = \lambda_j\phi_{j-}, \quad j=1,\ldots,N,
    \label{eq:constrFlowSourceAdjEigenProblem}
  \end{align}
\end{subequations}
where
\begin{align*}
  \left(\delta \lambda_j/\delta v,\,\delta \lambda_j/\delta p\right)^T
  &=\left(\shft{\phi_{j-}}{-1}\phi_{j+},\,\phi_{j-}\phi_{j+}\right)^T.
\end{align*}

Define
\begin{displaymath}
  \tilde{\Gamma}_n= \sum_{i=0}^{n} \left(
    \begin{array}{cc}
      a_i & b_i\\
      c_i & -a_i
    \end{array}\right)
  \lambda^{-i}+\sum_{j=1}^N\frac{1}{\lambda^n(\lambda-\lambda_j)}
  \left(
    \begin{array}{cc}
      -\shft{v}{1}\phi_{j-}\shft{\phi_{j+}}{1} 
      & \shft{\phi_{j-}}{1}\shft{\phi_{j+}}{1}\\
      -\shft{v}{1}\phi_{j-}\phi_{j+} 
      & \shft{v}{1}\phi_{j-}\shft{\phi_{j+}}{1}
    \end{array} \right).
\end{displaymath}
The Lax representation for \eqref{eq:ConstraintFlow} is
\cite{art:Zeng97}
\begin{equation}
  \label{eq:LaxConsFlow}
  \shft{\Psi}{-1}=U\Psi,\quad\mu\Psi=\tilde{\Gamma}_n\Psi
\end{equation}

Following the idea in the continuous case that by treating the
constrained flows of soliton equation as the stationary equations of
the soliton equation with self-consistent sources, we define Toda
lattice hierarchy with $N$ self-consistent sources as the following
system
\begin{subequations}
  \label{subeqs:TLHSCS}
  \begin{align}
    \left(
      \begin{array}{c}
        v \\ p
      \end{array}
    \right)_{t_n} &= J
    \left[
      \begin{array}{c}
        \frac{\delta H_n}{\delta v} 
        + \sum_{j=1}^N \frac{\delta\lambda_j}{\delta v}  \\
        \frac{\delta H_n}{\delta p} 
        + \sum_{j=1}^N \frac{\delta\lambda_j}{\delta p}
      \end{array}
    \right]=J \left[
      \begin{array}{c}
        \shft{a_{n+1}}{-1}/v+\sum_{j=1}^N\shft{\phi_{j-}}{-1}\phi_{j+}\\ 
        -\shft{b_{n+1}}{-1}+\sum_{j=1}^N\phi_{j-}\phi_{j+}
      \end{array}
    \right],\label{eq:TodaWithSources}\\
    L \phi_{j+} & = \lambda_j\phi_{j+}, 
    \label{eq:SourceEigenProblem}\\
    L^* \phi_{j-} & = \lambda_j\phi_{j-},\quad j=1,\ldots,N.
    \label{eq:SourceAdjEigenProblem}
  \end{align}
\end{subequations}
The Lax representation for \eqref{eq:TodaWithSources} can be obtained
from the Lax representation \eqref{eq:LaxConsFlow}
\begin{equation}
  \label{eq:LaxTLHSCS}
  \shft{\Psi}{-1}=U\Psi,\quad 
  -\Psi_{t_n}=\left(\lambda^n\tilde{\Gamma}_n+\tilde{\Delta}_n\right)\Psi,
\end{equation}
where $\tilde{\Delta}_n=\diag
(b_{n+1}-\sum_{j=1}^N\shft{\phi_{j-}}{1}\shft{\phi_{j+}}{1}+\delta,\delta)$,
with $\delta$ an arbitrary constant.

For $n=1$, we get the Toda lattice equation with $N$ self-consistent
sources (TLSCS)
\begin{subequations}
  \label{subeqs:TLSCS}
  \begin{align}
    &v_t=v\bigg(\shft{p}{-1}+
    \sum_{j=1}^N\shft{\phi_{j-}}{-1}\shft{\phi_{j+}}{-1}\bigg)
    -v\bigg(p+\sum_{j=1}^N\phi_{j-}\phi_{j+}\bigg), 
    \label{eq:1thTLWithSources-v}\\
    &p_t=v\bigg(1+\sum_{j=1}^N\shft{\phi_{j-}}{-1}\phi_{j+}\bigg)-
    \shft{v}{1}\bigg(1+\sum_{j=1}^N\phi_{j-}\shft{\phi_{j+}}{1}\bigg),
    \label{eq:1thTLWithSources-p}\\
    &L\phi_{j+} = \lambda_j\phi_{j+}, \\
    &L^* \phi_{j-} = \lambda_j\phi_{j-},\quad j=1,\ldots,N.
  \end{align}
\end{subequations}

The Lax representation for TLSCS is obtained from \eqref{eq:LaxTLHSCS}
by taking $n=1$, and $\delta=\lambda/2$. We prefer to rewrite it
equivalently in the following scalar form
\begin{subequations}
  \label{subeqs:1thTLSCS-LaxPair}
  \begin{align}
    L\psi &= \shft{v}{1}\shft{\psi}{1} + p\psi +\shft{\psi}{-1} =
    \lambda \psi \label{eq:1thTLSCSEigenProb}\\
    -\psi_t &= \shft{v}{1}\shft{\psi}{1} + \sum_{j=1}^N
    \frac{1}{\lambda-\lambda_j}\shft{v}{1}\phi_{j-}\left(\shft{\phi_{j+}}{1}\psi
      - \phi_{j+} \shft{\psi}{1}\right)\label{eq:1thTLSCStPart}\\
    L \phi_{j+} &
    = \lambda_j\phi_{j+}, 
    \label{eq:1thTLSCSSourceEigen}\\
    L^* \phi_{j-} &
    = \lambda_j\phi_{j-},\quad j=1,\ldots,N.
    \label{eq:1thTLSCSSourceAdjEigen}
  \end{align}
\end{subequations}

With the substitutions
\begin{equation}
  \label{eq:rel-vp-x}
  v:=\exp(\shft{x}{-1}-x),\quad
  p:= x_t - \sum_{j=1}^N\phi_{j+}\phi_{j-},
\end{equation}
equations (\ref{eq:1thTLWithSources-v}) and
(\ref{eq:1thTLWithSources-p}) can be written as
\begin{multline}
  \label{eq:TLSCS-ODE}
  x_{tt}=\exp\left(\shft{x}{-1}-x\right)
  \bigg(1+\sum_{j=1}^N\phi_{j+}\shft{\phi_{j-}}{-1}\bigg)\\
  -\exp\left(x-\shft{x}{1}\right)
  \bigg(1+\sum_{j=1}^N\shft{\phi_{j+}}{1}\phi_{j-}\bigg)
  +\sum_{j=1}^N\left(\phi_{j+}\phi_{j-}\right)_t.
\end{multline}

\section{The Darboux Transformations for TLSCS}
\label{sec:DTTLSCS}

\subsection{The forward DT with arbitrary functions of time}

Based on the Darboux transformation for Toda equation in
\cite{art:StahlhofenMatveev95,bk:MatveevSalle91}, we can find the
following theorem.
\begin{theorem}\label{thm:oneFDT}
  Given the solution $v$, $p$, $x$, $\phi_{i\pm}$ $(i=1,\ldots,N)$ for
  \eqref{subeqs:TLSCS} and \eqref{eq:TLSCS-ODE}, and eigenfunction
  $\psi$ for \eqref{eq:1thTLSCSEigenProb} and
  \eqref{eq:1thTLSCStPart}, let $f$ and $g$ be two independent
  eigenfunctions of \eqref{eq:1thTLSCSEigenProb} and
  \eqref{eq:1thTLSCStPart} with $\lambda=\mu$. Denote $h=f+\alpha(t)g$
  with the coefficient $\alpha(t)$ being an arbitrary differentiable
  function of $t$. Then the FDT is defined as follows
  {\allowdisplaybreaks
  \begin{subequations}
    \label{subeqs:1FDT}
    \begin{align}
      \psi[1] &=\psi-\frac{h}{\shft{h}{1}}\shft{\psi}{1},\\
      v[1] &=\shft{v}{1}\frac{\shft{h}{1}\shft{h}{-1}}{h^2},\\
      p[1]&=p-\frac{h}{\shft{h}{1}}+\frac{\shft{h}{-1}}{h},\\
      x[1]&=\shft{x}{1}+\frac{1}{2}\log\left(\frac{h}{\shft{h}{1}}\right)^2,\\
      \phi_{i+}[1]&=\phi_{i+}-\frac{h}{\shft{h}{1}}\shft{\phi_{i+}}{1},\\
      \phi_{i-}[1]&=\frac{\Delta^{-1}E\left(h\phi_{i-}\right)+\kappa_i(t)}{h},
      \quad i=1,\ldots,N
      \label{eq:phi_i-1FDT}\\
      \phi_{N+1,+}[1] &=c(t)\left(f-\frac{h}{\shft{h}{1}}\shft{f}{1}\right),\\
      \phi_{N+1,-}[1] &= d(t)\frac{1}{h},\label{eq:phi_N+1-1FDT}
    \end{align}
  \end{subequations}
}where $c(t)$ and $d(t)$ are arbitrary fixed differentiable functions
of $t$ satisfying $c(t)\cdot d(t)= -\dot{\alpha}/\alpha$,
$\kappa_i(t)=\frac{1}{\mu-\lambda_i}
[\shft{v}{1}\shft{h}{1}\phi_{i-}-h\shft{\phi_{i-}}{1}]|_{n=0}$. Namely
$v[1]$,$p[1]$, $x[1]$,$\phi_{i\pm}[1]$ $(i=1,\ldots,N+1)$,
$\lambda_{N+1}=\mu$ and $\psi[1]$ give a new solution to
\eqref{subeqs:TLSCS} or \eqref{eq:TLSCS-ODE} and
\eqref{subeqs:1thTLSCS-LaxPair} with $N+1$ self-consistent sources.
\end{theorem}

This theorem can be proved by straightforward calculation.

\begin{remark}
  Theorem~\ref{thm:oneFDT} serves as a non-auto-B\"{a}cklund
  transformation between two TLSCSs with degrees of
  sources $N$ and $N+1$
\end{remark}

We give an example for obtaining solution via Theorem
\ref{thm:oneFDT}.

\begin{example}[Rational solutions]
  \label{ex:rationalSolution}
  Starting from trivial solution $v=1$,$p=0$ and vanishing sources,
  the Lax pair \eqref{subeqs:1thTLSCS-LaxPair} becomes
  \begin{subequations}
    \label{subeqs:trivialLaxPair}
    \begin{align}
      L\psi &= \shft{\psi}{1} + \shft{\psi}{-1} = \lambda \psi,
      \label{eq:1thTLEigenProb}\\
      -\psi_t& = \shft{\psi}{1} .
      \label{eq:1thTLtPart}
    \end{align}
  \end{subequations}
  Let $\psi=\exp(kn-e^kt)$ be the solution of
  \eqref{subeqs:trivialLaxPair} w.r.t. $\lambda = 2\cosh(k)$, $f =
  (an-at+b)\exp(-t)$, $g = \exp(-t)$ be independent solutions of
  \eqref{subeqs:trivialLaxPair} w.r.t. $\lambda=2$, where
  $a\in\mathbb{R}-\{0\}$ and $b\in\mathbb{R}$ are arbitrary constants.
  Let $h = f + \alpha(t)g$ with the differentiable function
  $\alpha(t)$. Then we obtain the rational solution with a pair of
  non-vanishing sources{\allowdisplaybreaks
    \begin{align*}
      \psi_+[1] &=
      \left[1-e^k\frac{an+b-at+\alpha(t)}{an+a+b-at+\alpha(t)}\right]
      \exp(kn-e^kt), \\
      v[1] &=
      \frac{\left[an+a+b-at+\alpha(t)\right]\left[an-a+b-at+\alpha(t)\right]}
      {\left[an+b-at+\alpha(t)\right]^2},\\
      p[1] &= \frac{an-a+b-at+\alpha(t)}{an+b-at+\alpha(t)}
      -\frac{an+b-at+\alpha(t)}{an+a+b-at+\alpha(t)},\\
      x[1] &=
      \frac{1}{2}\log\left[\frac{an+b-at+\alpha(t)}{an+a+b-at+\alpha(t)}\right]^2,\\
      \phi_{1+}[1] &= -c(t)\frac{ae^{-t}\alpha(t)}{an+a+b-at+\alpha(t)},\\
      \phi_{1-}[1] &= d(t)\frac{e^t}{an+b-at+\alpha(t)},
    \end{align*}
  } where $c(t)$ and $d(t)$ satisfying $c(t)d(t)=-\dot{\alpha}/\alpha$.
\end{example}

\subsection{The Multi-time Repeated FDT}

\begin{theorem}[The multi-time repeated FDT]
  \label{thm:N-timeDT}
  Given the solution $v$, $p$, $x$, $\phi_{i\pm}$ $(i=1,\ldots,N)$ for
  \eqref{subeqs:TLSCS} and \eqref{eq:TLSCS-ODE}, and eigenfunction
  $\psi$ for \eqref{eq:1thTLSCSEigenProb} and
  \eqref{eq:1thTLSCStPart}, let $f_j$ and $g_j$ be independent
  eigenfunctions of~\eqref{eq:1thTLSCSEigenProb} and
  \eqref{eq:1thTLSCStPart} w.r.t. distinct $\mu_j$ $(j=1,\ldots,l)$.
  Let $\alpha_j(t)$ $(j=1,\ldots,l)$ be arbitrary smooth functions of
  $t$. Denote $h_j=f_j+\alpha_jg_j$. Then the $l$-times repeated FDT
  is given as {\allowdisplaybreaks
    \begin{subequations}
      \label{subeqs:lFDTs}
      \begin{align}
        \psi[l] &=
        \frac{\cas{\psi,h_1,\ldots,h_l}}{\shft{\cas{h_1,\ldots,h_l}}{1}},
        \label{eq:DT_psi+Ntime}\\
        v[l] &= \shft{v}{l}\frac{\shft{\cas{h_1,\ldots,h_l}}{1}
          \shft{\cas{h_1,\ldots,h_l}}{-1}} {\cas{h_1,\ldots,h_l}^2},
        \label{eq:DT_v_Ntime}\\
        p[l] &= p+
        \frac{\shft{\wcas{h_1,\ldots,h_l}}{-1}}{\cas{h_1,\ldots,h_l}}
        -\frac{\wcas{h_1,\ldots,h_l}}{\shft{\cas{h_1,\ldots,h_l}}{1}},
        \label{eq:DT_p_Ntime}\\
        x[l] &= \shft{x}{l}+\frac{1}{2}\log\left[
          \frac{\cas{h_1,\ldots,h_l}}{\shft{\cas{h_1,\ldots,h_l}}{1}}\right]^2,
        \label{eq:DT_x_Ntime}\\
        \phi_{i+}[l] &= \frac{\cas{\phi_{i+},h_1,\ldots,h_l}}
        {\shft{\cas{h_1,\ldots,h_l}}{1}},
        \label{eq:DT_old_phi_i+_Ntime}\\
        \phi_{i-}[l] &= \frac{\bcas{\phi_{i-},h_1,\ldots,h_l}}
        {\cas{h_1,\ldots,h_l}},\quad i=1,\ldots,N
        \label{eq:DT_old_phi_i-_Ntime}\\
        \phi_{N+j,+}[l] &= c_j(t)
        \frac{\cas{f_j,h_1,\ldots,h_l}}{\shft{\cas{h_1,\ldots,h_l}}{1}},
        \label{eq:DT_phi_j+_Ntime}\\
        \phi_{N+j,-}[l] &= d_j(t)
        \frac{\shft{\cas{h_1,\ldots,\widehat{h_j},\ldots,h_l}}{1}}{\cas{h_1,\ldots,h_l}},
        \quad j=1,\ldots,l
        \label{eq:DT_phi_j-Ntime}
      \end{align}
    \end{subequations}
  } where 
  \begin{displaymath}
    \cas{h_1,\ldots,h_l}:=
    \det\left(\shft{h_i}{j}\right)_{\substack{i=1,\ldots,l\\ j=0,\ldots,l-1}},
  \end{displaymath}
  is the Casorati determinant, and
  \begin{displaymath}
    \wcas{h_1,\ldots,h_l}:=
    \det\left(\shft{h_i}{j}\right)_{\substack{i=1,\ldots,l \\ j=0,2,\ldots,l}},
  \end{displaymath}
  \begin{displaymath}
    \bcas{\phi_{i-},h_1,\ldots,h_l}:=\det\left[
      \begin{array}{ccc}
        S_{i}(h_1)&\cdots&S_{i}(h_l)\\
        \shft{h_1}{1}&\cdots&\shft{h_l}{1}\\
        \vdots&\vdots&\vdots\\
        \shft{h_1}{l-1}&\cdots&\shft{h_l}{l-1}
      \end{array}\right]
  \end{displaymath}
  where 
  \begin{displaymath}
    S_{i}(h_j):=S(\phi_{i-},h_j)=\Delta^{-1}E(\phi_{i-}h_j)+
    (\mu_j-\lambda_i)^{-1}[\shft{v}{1}\shft{h_j}{1}\phi_{i-}-h_j\shft{\phi_{i-}}{1}]|_{n=0}.
  \end{displaymath}
  The $\widehat{h_j}$ means the removal of this term and $c_j(t)\cdot
  d_j(t)=(-1)^j\dot{\alpha_j}/\alpha_j$. Functions $\psi[l]$, $v[l]$,
  $p[l]$, $x[l]$, $\phi_{i\pm}[l]$ $(i=1,\ldots,N+l)$ and
  $\lambda_{N+j}=\mu_j$ $(j=1,\ldots,l)$ satisfy
  \eqref{subeqs:1thTLSCS-LaxPair},\eqref{subeqs:TLSCS} and
  \eqref{eq:TLSCS-ODE} with $N$ replaced by $N+l$.
\end{theorem}

\begin{proof}[Sketch of Proof.]

  For the proof of formulae (\ref{eq:DT_psi+Ntime}),
  (\ref{eq:DT_v_Ntime}), (\ref{eq:DT_p_Ntime}),
  (\ref{eq:DT_old_phi_i+_Ntime}), (\ref{eq:DT_phi_j+_Ntime}) see
  \cite{bk:MatveevSalle91}. The formula \eqref{eq:DT_x_Ntime} is
  proved by using~\eqref{eq:rel-vp-x}. Formulae
  \eqref{eq:DT_old_phi_i-_Ntime} and \eqref{eq:DT_phi_j-Ntime} are
  proved by induction. The calculation is lengthy but rather straight
  forward. We omit it.

\end{proof}
\begin{remark}
  The multi-times repeated FDT \eqref{subeqs:lFDTs} provides a
  non-auto-B\"acklund transformation between two TLSCSs with degrees
  $N$ and $N+l$.
\end{remark}

\subsection{The generalized forward Darboux transformations}
It is well known from \cite{art:StahlhofenMatveev95} that the positon
solution of the Toda lattice equations are obtained by computing the
limit $k_2\rightarrow k_1$ in the result of two-step DT, where
$k_{1,2}$ are parameters of eigenfunction with which DT is generated.
In order to construct positon solutions for TLSCS, similar
consideration can be made in our case of FDT with arbitrary functions
of time. However, in our case, the arbitrary time functions
$\alpha_j(t)$ must be carefully chosen to balance the divergence of
sources. Possible candidates for $\alpha_j(t)$ are exponential
functions. By using it, we get the following GFDT.
\begin{theorem}[GFDT]
  \label{thm:deGFDT}
  Given the solution $v$, $p$, $x$, $\phi_{i\pm}$ $(i=1,\ldots,N)$ for
  \eqref{subeqs:TLSCS} and \eqref{eq:TLSCS-ODE}, and eigenfunction
  $\psi$ for \eqref{eq:1thTLSCSEigenProb} and
  \eqref{eq:1thTLSCStPart}, let $F_j$, $G_j$ $(j=1,\ldots,l)$ be pairs
  of independent eigenfunctions of \eqref{eq:1thTLSCSEigenProb} and
  \eqref{eq:1thTLSCStPart} corresponding to distinct $\lambda_{N+j}$.
  Let $f_r$, $g_r$ $(r=1,\ldots,I)$ be pairs of independent
  eigenfunctions of \eqref{eq:1thTLSCSEigenProb} and
  \eqref{eq:1thTLSCStPart} corresponding to
  $\lambda_{N+l+r}=\lambda_{N+l+r}(\omega_r)$, where
  $\lambda_{N+l+r}(\omega_r)$ is analytic function of parameter
  $\omega_r\in\mathbb{C}$. Let $\alpha_j(t)$ $(j=1,\ldots, l)$,
  $\beta_r(t)$ $(r=1,\ldots,I)$ be arbitrary differentiable functions
  of $t$. Let $m_r\in\mathbb{N},m_r\geq 2$ $(r=1,\ldots, I)$,
  $\mathbf{m}:=(m_1,\ldots,m_I)$. Denote $q_j=F_j+\alpha_j(t)G_j$,
  $h_r=f_r+g_r$ respectively. Then the following GFDT with $l$ times
  of FDT with arbitrary time functions $\alpha_j(t)$ and $I$ times of
  generalized FDT of multiplicities $m_r$ with arbitrary time
  functions $\beta_r(t)$ is given by 
  {\allowdisplaybreaks
    \begin{subequations}
      \begin{align}
        \psi[l,\mathbf{m}] &=
        \frac{\Delta_{l,\mathbf{m}}(\psi)}{\shft{\Delta_{l,\mathbf{m}}}{1}},
        \label{eq:psi-deGFDT}\\
        v[l,\mathbf{m}] &=\shft{v}{l+|\mathbf{m}|}
        \frac{\shft{\Delta_{l,\mathbf{m}}}{1}\shft{\Delta_{l,\mathbf{m}}}{-1}}{\Delta_{l,\mathbf{m}}^2},
        \label{eq:v-deGFDT}\\
        p[l,\mathbf{m}] &=p+
        \frac{\shft{\widetilde{\Delta}_{l,\mathbf{m}}}{-1}}{\Delta_{l,\mathbf{m}}}-
        \frac{\widetilde{\Delta}_{l,\mathbf{m}}}{\shft{\Delta_{l,\mathbf{m}}}{1}},
        \label{eq:p-deGFDT}\\
        x[l,\mathbf{m}] &=\shft{x}{l+|\mathbf{m}|}+
        \frac{1}{2}\log\left(\frac{\Delta_{l,\mathbf{m}}}{\shft{\Delta_{l,\mathbf{m}}}{1}}\right)^2,
        \label{eq:x-deGFDT}\\
        \phi_{i+}[l,\mathbf{m}] &=
        \frac{\Delta_{l,\mathbf{m}}(\phi_{i+})}{\shft{\Delta_{l,\mathbf{m}}}{1}}, &
        \label{eq:oldphi+deGFDT}\\
        \phi_{i-}[l,\mathbf{m}] &=
        \frac{\overline{\Delta}_{l,\mathbf{m}}(\phi_{i-})}{\Delta_{l,m}}, &
        i=1,\ldots,N
        \label{eq:oldphi-deGFDT}\\
        \phi_{N+j,+}[l,\mathbf{m}] &= 
        c_j(t)\frac{\Delta_{l,\mathbf{m}}(F_j)}{\shft{\Delta_{l,\mathbf{m}}}{1}}, 
        \label{eq:phil+deGFDT}\\
        \phi_{N+j,-}[l,\mathbf{m}] &= 
        d_j(t)\frac{\shft{{\Delta_{l,\mathbf{m}}^j}}{1}}{\Delta_{l,\mathbf{m}}}, & 
        j=1,\ldots,l
        \label{eq:phil-deGFDT}\\
        \phi_{N+l+r,+}[l,\mathbf{m}] &=
        -\frac{\Delta_{l,\mathbf{m}}(f_r)}{\shft{\Delta_{l,\mathbf{m}}}{1}}, & 
        \label{eq:phin+deGFDT}\\
        \phi_{N+l+r,-}[l,\mathbf{m}] &=\dot{\beta_r}
        \frac{\shft{{\Delta_{l,\mathbf{m}}^{l+r}}}{1}}{\Delta_{l,\mathbf{m}}}, & 
        r=1,\ldots,I
        \label{eq:phin-deGFDT}
      \end{align}
    \end{subequations}
  } where $c_j(t)\cdot d_j(t)=(-1)^j\dot{\alpha_j}/\alpha_j$,
  $|\mathbf{m}|:=\sum m_r$. Symbols are defined as follows:
  {\allowdisplaybreaks
    \begin{subequations}
      \begin{align*}
        \Delta_{l,\mathbf{m}}(\psi) := \hdcas(&\psi,q_1,q_2,\ldots,q_l; \\
        & h_1,\partial_{\omega_1} h_1,\partial_{\omega_1}^2h_1,
        \ldots,
        \partial_{\omega_1}^{m_1-1}h_1+(-1)^{m_1-1}\beta_1(t)g_1;\cdots;\\
        & h_I,\partial_{\omega_I} h_I,\partial_{\omega_I}^2 h_I
        ,\ldots,
        \partial_{\omega_I}^{m_I-1}h_I+(-1)^{m_I-1}\beta_I(t)g_I),
      \end{align*}
      \begin{align*}
        \Delta_{l,\mathbf{m}} := \hdcas(&q_1,q_2,\ldots,q_l; \\
        & h_1,\partial_{\omega_1} h_1,\partial_{\omega_1}^2h_1
        ,\ldots,
        \partial_{\omega_1}^{m_1-1}h_1+(-1)^{m_1-1}\beta_1(t)g_1;\cdots;\\
        & h_I,\partial_{\omega_I} h_I,\partial_{\omega_I}^2 h_I
        ,\ldots,
        \partial_{\omega_I}^{m_I-1}h_I+(-1)^{m_I-1}\beta_I(t)g_I),
      \end{align*}
      \begin{align*}
        \widetilde{\Delta}_{l,\mathbf{m}} := \hdwcas(&q_1,q_2,\ldots,q_l; \\
        & h_1,\partial_{\omega_1} h_1,\partial_{\omega_1}^2h_1
        ,\ldots,
        \partial_{\omega_1}^{m_1-1}h_1+(-1)^{m_1-1}\beta_1(t)g_1;\cdots;\\
        & h_I,\partial_{\omega_I} h_I,\partial_{\omega_I}^2 h_I
        ,\ldots,
        \partial_{\omega_I}^{m_I-1}h_I+(-1)^{m_I-1}\beta_I(t)g_I),
      \end{align*}
      \begin{align*}
        \Delta_{l,\mathbf{m}}^j :=\hdcas(&q_1,\ldots,\widehat{q_j},\ldots,q_l; \\
        & h_1,\partial_{\omega_1} h_1,\partial_{\omega_1}^2h_1
        ,\ldots,
        \partial_{\omega_1}^{m_1-1}h_1+(-1)^{m_1-1}\beta_1(t)g_1;\cdots;\\
        & h_I,\partial_{\omega_I} h_I,\partial_{\omega_I}^2 h_I
        ,\ldots,
        \partial_{\omega_I}^{m_I-1}h_I+(-1)^{m_I-1}\beta_I(t)g_I)\quad(1\leq
        j\leq l),
      \end{align*}
      \begin{align*}
        \Delta_{l,\mathbf{m}}^{l+r} := \hdcas(&q_1,q_2,\ldots,q_l; \\
        & h_1,\partial_{\omega_1} h_1,\partial_{\omega_1}^2h_1
        ,\ldots,
        \partial_{\omega_1}^{m_1}h_1+(-1)^{m_1-1}\beta_1(t)g_1;\cdots;\\
        & h_r,\partial_{\omega_r} h_r,\partial_{\omega_r}^2h_r
        ,\ldots,
        \partial_{\omega_r}^{m_r-2}h_r;\cdots;\\
        &h_I,\partial_{\omega_I} h_I,\partial_{\omega_I}^2 h_I
        ,\ldots,
        \partial_{\omega_I}^{m_I-1}h_I+(-1)^{m_I-1}\beta_I(t)g_I)\quad(1\leq
        r\leq I).
      \end{align*}
      where $\hdcas$ and $\hdwcas$ are defined in Theorem
      \ref{thm:N-timeDT}.
      \begin{multline*}
         \overline{\Delta}_{l,\mathbf{m}}(\phi_{i-}):=
        \det\Big(\mathbf{w}_i(q_1),\ldots,\mathbf{w}_i(q_1);\\
         \mathbf{w}_i(h_1),\partial_{\omega_1}\mathbf{w}_i(h_1),\partial_{\omega_1}^2\mathbf{w}_i(h_1),
        \ldots,\partial_{\omega_1}^{m_1-1}\mathbf{w}_i(h_1)+(-1)^{m_1-1}\beta_1\mathbf{w}_i(g_1);\cdots;\\
         \mathbf{w}_i(h_I),\partial_{\omega_I}\mathbf{w}_i(h_I),\partial_{\omega_I}^2\mathbf{w}_i(h_I),
        \ldots,\partial_{\omega_I}^{m_I-1}\mathbf{w}_i(h_I)+(-1)^{m_I-1}\beta_I\mathbf{w}_i(g_I)\Big),
      \end{multline*}
      where for any solution $\psi$ of \eqref{eq:1thTLSCSEigenProb}
      with eigenvalue $\lambda$, $\mathbf{w}_i(\psi)$ is
      $l+|\mathbf{m}|$ dimensional column vector defined as
      \begin{displaymath}
        \mathbf{w}_i(\psi)=(S_i(\psi),\shft{\psi}{1},\ldots,\shft{\psi}{l+|\mathbf{m}|-1})^T,
      \end{displaymath}
      and $S_i(\psi)$ is a scalar defined as
      \begin{displaymath}
        S_i(\psi):=S(\phi_{i-},\psi)=\Delta^{-1}E(\phi_{i-}\psi)+
        (\lambda-\lambda_i)^{-1}[\shft{v}{1}\shft{\psi}{1}\phi_{i-}-\phi_{i-}\shft{\psi}{1}]|_{n=0}.
      \end{displaymath}
    \end{subequations}
    Then $\psi[l,\mathbf{m}]$, $v[l,\mathbf{m}]$, $p[l,\mathbf{m}]$,
    $x[l,\mathbf{m}]$, $\phi_{i\pm}[l,\mathbf{m}]$ and $\lambda_i$
    $(i=1,\ldots,N+l+I)$ satisfy
    \eqref{subeqs:1thTLSCS-LaxPair},\eqref{subeqs:TLSCS} and
    \eqref{eq:TLSCS-ODE} with $N$ replaced by $N+l+I$. }
\end{theorem}
\begin{proof}[Proof.]
  Without loss of generality, we only prove the special case $l=0$,
  $I=1$. The multiplicity $m_1$ is denoted by $m$ ($m\geq 2$). And
  $\lambda_1(\omega_1)$ is denoted by $\lambda(\omega)$, which is
  analytic function of parameter $\omega$. $f_1$, $g_1$, $h_1$ and
  $\beta_1(t)$ are denoted by $f$, $g$, $h$ and $\beta(t)$
  respectively.

  Let $\omega_s=\omega + \varepsilon e_s$, where $e_s$
  ($s=1,\ldots,m$) are distinct complex constants, $\varepsilon$ is a
  small parameter. Define $\varrho_s(t)=\exp(\Omega_sb_s(t))$ , where
  $b_s(t)$ are arbitrary differentiable functions of $t$ satisfying
  $\sum_{s=1}^m b_s(t) = \beta(t)$, and
  \begin{displaymath}
    \Omega_s = \frac{1}{(m-1)!}
    \prod_{\substack{1\leq i \leq m \\ i \neq s}}(\omega_i - \omega_s),\quad
    p_s = \frac{1}{(m-1)!}\prod_{\substack{1\leq i \leq m \\ i \neq s}}(e_i - e_s).
  \end{displaymath}
  We have the obvious relation $\Omega_s = \varepsilon^{m-1}p_s$.
  
  We have the following important observations. Let
  $\mathbf{u}(\lambda)$, $\mathbf{v}(\lambda)$ be two column vectors
  of dimension $m$, $\bar{\mathbf{u}}(\lambda)$,
  $\bar{\mathbf{v}}(\lambda)$ be two column vectors of dimension
  $m-1$, whose components are analytic functions of
  $\lambda=\lambda(\omega)$. Denote
  $\mathbf{u}_s=\mathbf{u}(\lambda(\omega_s))$,
  $\mathbf{v}_s=\mathbf{v}(\lambda(\omega_s))$,
  $\bar{\mathbf{u}}_s=\bar{\mathbf{u}}(\lambda(\omega_s))$,
  $\bar{\mathbf{v}}_s=\bar{\mathbf{v}}(\lambda(\omega_s))$ then
  \begin{align}
    &\det(\mathbf{u}_1+\varrho_1 \mathbf{v}_1,\ldots, 
    \mathbf{u}_m+\varrho_m \mathbf{v}_m )\notag\\
    =&\frac{\varepsilon^{\frac{1}{2}m(m-1)}}{1!\cdots(m-1)!}
    \prod_{1\leq i<j \leq m}(e_j-e_i)
    \det\Big(\mathbf{u}+\mathbf{v},
    \partial_\omega (\mathbf{u}+\mathbf{v}),\notag\\
    &\ldots,\partial_\omega^{m-1}(\mathbf{u}+\mathbf{v})
    +(-1)^{m-1}\beta(t)\mathbf{v}\Big)+o(\varepsilon^{\frac{1}{2}m(m-1)})
    \label{eq:det-expension1},\\
    &\det(\bar{\mathbf{u}}_1+\varrho_1 \bar{\mathbf{v}}_1,\ldots,
    \widehat{\bar{\mathbf{u}}_r+\varrho_r\bar{\mathbf{v}}_r},\ldots, 
    \bar{\mathbf{u}}_m+\varrho_m \bar{\mathbf{v}}_m )\notag\\
    =&\frac{\varepsilon^{\frac{1}{2}(m-2)(m-1)}}{1!\ldots(m-2)!}
    \prod_{\substack{1\leq i<j\leq m \\ i, j\neq r}}(e_j-e_i)
    \det\Big(\bar{\mathbf{u}}+\bar{\mathbf{v}},
      \partial_\omega (\bar{\mathbf{u}}+\bar{\mathbf{v}}),\ldots,
      \partial_\omega^{m-2}(\bar{\mathbf{u}}+\bar{\mathbf{v}})\Big)\notag\\
    &+o(\varepsilon^{\frac{1}{2}(m-2)(m-1)})\label{eq:det-expension2},
  \end{align}

  These observations can be easily obtained by insert the Taylor
  expansion of $\mathbf{u}_s$, $\mathbf{v}_s$ and
  $\bar{\mathbf{u}}_s$, $\bar{\mathbf{v}}_s$ at $\varepsilon = 0$ into
  the determinant.

  By applying multi-time repeated FDT (Theorem \ref{thm:N-timeDT})
  with $h(\lambda(\omega_s))=f(\lambda(\omega_s))+\varrho_s(t)
  g(\lambda(\omega_s))$, $c_s=-1$,
  $d_s=(-1)^{s-1}\dot{\varrho_s}/\varrho_s$ $(s=1,\ldots,m)$, we
  obtain $\psi[m]$, $v[m]$, $p[m]$, $x[m]$, $\phi_{i\pm}[m]$
  $(i=1,\ldots,N+m)$ and $\lambda_{N+s}=\lambda(\omega_s)$ satisfying
  \eqref{subeqs:1thTLSCS-LaxPair}, \eqref{subeqs:TLSCS} and
  \eqref{eq:TLSCS-ODE} with $N+m$ self-consistent sources. Since such
  solution contain an arbitrary parameter $\varepsilon$, letting
  $\varepsilon \rightarrow 0$ will also give rise to solution to
  \eqref{subeqs:1thTLSCS-LaxPair}, \eqref{subeqs:TLSCS} and
  \eqref{eq:TLSCS-ODE} with $N+m$ self-consistent sources. So by doing
  this with use of \eqref{eq:det-expension1} and
  \eqref{eq:det-expension2}, we get the special case
  $(l=0,\mathbf{m}=m)$ of formulae \eqref{eq:psi-deGFDT},
  \eqref{eq:v-deGFDT}, \eqref{eq:p-deGFDT}, \eqref{eq:x-deGFDT},
  \eqref{eq:oldphi+deGFDT} and \eqref{eq:oldphi-deGFDT}.  We also
  obtain $m$ self-consistent sources
  \begin{align*}
    \phi_{N+s,+}[m] &= 
    -\frac{\cas{f,h,\ldots,\partial_\omega^{m-1}h}}
    {\cas{h,\ldots,\partial_\omega^{m-1}h+(-1)^{m-1}\beta(t)g}},\\
    \phi_{N+s,-}[m] &= \dot{b_s}\frac{\cas{h,\ldots,\partial_\omega^{m-2}h}}
    {\cas{h,\ldots,\partial_\omega^{m-1}h+(-1)^{m-1}\beta(t)g}},
    \quad s=1,\ldots,m.
  \end{align*}
  Since $\phi_{N+s,+}[m]$'s are equal,
  $\phi_{N+s,-}[m]$ differ only in coefficients $\dot{b_s}$ and they
  all corresponding to eigenvalue $\lambda(\omega)$, it is
  reasonable to combine $m$ self-consistent sources to one
  self-consistent source by denoting
  \begin{align*}
    \phi_{N+1,+}[0,m] &= \phi_{N+1,+}[m],\\
    \phi_{N+1,-}[0,m] &= \sum_{s=1}^m\phi_{N+s,-}[m].
  \end{align*}
  Thus we arrive at \eqref{eq:phin+deGFDT} and \eqref{eq:phin-deGFDT}.
  
  The general case can be proved by repeating above procedure.
\end{proof}

\section{Solutions of TLSCS}
\label{sec:solut-toda-latt}

The GFDT technique enable us to construct various types of solutions
to the TLSCS. Starting from trivial solution $v=1$, $p=0$ for
\eqref{subeqs:TLSCS} with $N=0$, by choosing specific solutions of Lax
pair \eqref{subeqs:trivialLaxPair} and specific $l$, $I$ and
$\mathbf{m}$ in Theorem \ref{thm:deGFDT}, we can construct
multi-solitons solutions, (multi-)positon solutions (or higher order),
(multi-)negaton solutions (or higher order) and
(multi-)soliton-positon solutions, (multi-)soliton-negaton solutions,
(multi-)positon-negaton solutions etc..

\subsection{Soliton solutions }
\label{sec:oneSoliton}
Let $F = \nu^n\exp(n\gamma-\nu e^\gamma t)$,$G =
\nu^n\exp(-n\gamma-\nu e^{-\gamma}t)$ be solutions of
\eqref{subeqs:trivialLaxPair} with respect to
$\lambda=2\nu\cosh(\gamma)$, where
$\gamma\in\mathbb{R}\backslash\{0\}$. Let $\alpha=\exp(-2a(t))$, where
$a(t)$ is an arbitrary differentiable functions of $t$, $\nu=\pm 1$ .
Define
\begin{displaymath}
  q = F + \alpha G = 2\nu^n\exp(-\nu\cosh(\gamma)t-a)\cosh(U), \quad
  U := n\gamma-\nu\sinh(\gamma)t+a.
\end{displaymath}
Then using theorem \ref{thm:deGFDT} with $l=1$, $I=0$, we have the
1-soliton solution for \eqref{subeqs:TLSCS} or \eqref{eq:TLSCS-ODE}
with $N=1$.
{\allowdisplaybreaks
  \begin{align*}
    \psi^{\textrm{sol}}&=\exp(ikn-e^{ik}t)\left[1-\nu e^{ik}
      \frac{\cosh(U)}{\cosh(U+\gamma)}\right],\\
    v^{\textrm{sol}}&=\frac{\cosh(U-\gamma)\cosh(U+\gamma)}{\cosh^2(U)},\\
    p^{\textrm{sol}}&=\nu\frac{\cosh(U-\gamma)}{\cosh(U)}-\nu\frac{\cosh(U)}{\cosh(U+\gamma)},\\
    x^{\textrm{sol}}&=\log\left[\frac{\cosh(U)}{\cosh(U+\gamma)}\right],\\
    \phi_+^{\textrm{sol}}&=-\nu^n\sinh(\gamma)c(t)\frac{\exp(-\nu\cosh(\gamma)t-a)}{\cosh(U+\gamma)},\\
    \phi_-^{\textrm{sol}}&=\nu^nd(t)\frac{\exp(\nu\cosh(\gamma)t+a)}{2\cosh(U)},
  \end{align*}
} where $c(t)d(t)=2\dot a$. 

The 1-soliton solution of TLSCS is similar to the 1-soliton of
ordinary Toda lattice equation~\cite{bk:TodaLatt81}. The main
difference between them is that the travelling speed of 1-soliton
solution for TLSCS can vary with $t$, for the travelling speed is
$\frac{\nu\sinh\gamma-\dot{a}}{\gamma}$. The multi-soliton solutions
for TLSCS \eqref{subeqs:TLSCS} can be constructed similarly. We omit
it.




\subsection{Positon solutions and negaton solutions}
\label{sec:one-positon-solution}
\subsubsection{One-positon solution}
Let$f=c_1\exp(-t\cos\omega)\cos(n\omega-t\sin\omega)$,
$g=c_2\exp(-t\cos\omega)\sin(n\omega-t\sin\omega)$ be solutions of
\eqref{subeqs:trivialLaxPair} w.r.t $\lambda=2\cos\omega$. Set
constants $c_1=\cos\theta$, $c_2=\sin\theta$, where
$\theta\neq\frac{n}{2}\pi$ ($n\in \mathbb{Z}$). Define
\begin{displaymath}
  h=f+g=\exp(-t\cos\omega)\cos(Y), \quad
  Y=n\omega-t\sin\omega-\theta,
\end{displaymath}
and
\begin{displaymath}
  Z=\partial_\omega Y = n-t\cos\omega, \quad
  \eta=Z+\frac{1}{2}\beta(t)\sin(2\theta).
\end{displaymath}
where $\beta(t)$ is an arbitrary differentiable function. Then using
theorem~\ref{thm:deGFDT} with the specification $l=0$, $I=1$, $m_1=2$,
one gets one-positon solution for TLSCS \eqref{subeqs:TLSCS} with
$N=1$.
{\allowdisplaybreaks
  \begin{subequations}
    \begin{align}
      v^{\textrm{pos}}&=\frac
      {\left[(\eta+\frac{3}{2})\sin\omega+\frac{1}{2}\sin(2Y+3\omega)\right]
        \left[(\eta-\frac{1}{2})\sin\omega+\frac{1}{2}\sin(2Y-\omega)\right]}
      {\left[(\eta+\frac{1}{2})\sin\omega+\frac{1}{2}\sin(2Y+\omega)\right]^2}\\
      p^{\textrm{pos}}&=\frac
      {\eta\sin(2\omega)+\sin(2Y)}{(\eta+\frac{1}{2})\sin\omega+\frac{1}{2}\sin(2Y+\omega)}
      -\frac {(\eta+1)\sin(2\omega)+\sin(2Y+2\omega)}
      {(\eta+\frac{3}{2})\sin\omega+\frac{1}{2}\sin(2Y+3\omega)}\\
      x^{\textrm{pos}}&=\frac{1}{2}\log\left[
        \frac{(\eta+\frac{1}{2})\sin\omega+\frac{1}{2}\sin(2Y+\omega)}
        {(\eta+\frac{3}{2})\sin\omega+\frac{1}{2}\sin(2Y+3\omega)}
      \right]^2\label{eq:positon_x}\\
      \phi_+^{\textrm{pos}}&=-\sin(2\theta)
      \frac{\exp(-t\cos\omega)\sin^2\omega\cos(Y+\omega)}
      {(\eta+\frac{3}{2})\sin\omega+\frac{1}{2}\sin(2Y+3\omega)}
       \label{eq:positon_psource}\\
      \phi_-^{\textrm{pos}}&=-\dot{\beta}
      \frac{\exp(t\cos\omega)\cos(Y+\omega)}
      {(\eta+\frac{1}{2})\sin\omega+\frac{1}{2}\sin(2Y+\omega)}
      \label{eq:positon_msource}
    \end{align}
  \end{subequations}
}
The scattering properties can be analyzed, resembling with
\cite{art:StahlhofenMatveev95}.
It is noticed that the scattering matrix is an identity matrix, which
is the basic feature for positon solutions.

Then we discuss the analytic properties of one-positon solution. An
investigation on positon profile $x^{\textrm{pos}}$ shows that the
one-positon profile decays to zero slowly at $\pm\infty$, oscillates
and possesses singularities during their propagation. The
singularities of one positon profile are determined by zeros of
\begin{displaymath}
  \Delta^{\textrm{pos}}:=(\eta+\frac{1}{2})\sin\omega+\frac{1}{2}\sin(2Y+\omega).
\end{displaymath}

The speed of positon profile are defined by the speed of singularities
\cite{art:Matveev2002}. Assuming $n_0(t)$ to be one of the places
where singularity occurs, differentiating by $t$ on the two side of
the equation$\Delta^{\textrm{pos}}(n_o,t)=0$, one finds that the
singularity propagation is governed by a nonlinear ODE
\begin{displaymath}
  \dot{n_0} = \frac{\sin\omega}{\omega}\left[
    1+\frac{\omega\cos\omega-\frac{1}{2}\omega\dot{\beta}\sin(2\theta)-\sin\omega}
    {\sin\omega+\omega\cos(2n_0\omega-2t\sin\omega-2\theta+\omega)}\right].
\end{displaymath}

The singularities of $\phi_+^{\textrm{pos}}$ and
$\phi_-^{\textrm{pos}}$ are determined by zeros of
$\shft{(\Delta^{\textrm{pos}})}{1}$ and $\Delta^{\textrm{pos}}$,
respectively.


\subsubsection{Two-positon solution}
\label{sec:two-positon-solution}

Let
$f_i:=\cos\theta_i\exp(-t\cos\omega_i)\cos(n\omega_i-t\sin\omega_i)$,
$g_i:=\sin\theta_i\exp(-t\cos\omega_i)\sin(n\omega_i-t\sin\omega_i)$,
$i=1,2$ be solutions of \eqref{subeqs:trivialLaxPair} w.r.t
$\lambda_i=2\cos\omega_i$, $\theta_i\neq\frac{n}{2}\pi$ ($n\in
\mathbb{Z}$). Define
\begin{displaymath}
  h_i=f_i+g_i=\exp(-t\cos\omega_i)\cos(Y_i), \quad
  Y_i=n\omega_i-t\sin\omega_i-\theta_i,
\end{displaymath}
\begin{displaymath}
  Z_i=\partial_{\omega_i} Y_i = n-t\cos\omega_i, \quad
  \eta_i=Z_i+\frac{1}{2}\beta_i(t)\sin(2\theta_i).
\end{displaymath}
Then using theorem~\ref{thm:deGFDT} with $l=0$, $I=2$,
$\mathbf{m}=(2,2)$, one obtains the 2-positon solution for TLSCS
\eqref{subeqs:TLSCS} ,

\begin{align*}
  x^{\textrm{2p}}&=\frac{1}{2}\log\left[
    \frac{\cas{\cos Y_1,\eta_1\sin Y_1,\cos Y_2, \eta_2\sin Y_2}}
    {\shft{\cas{\cos Y_1,\eta_1\sin Y_1, \cos Y_2, \eta_2\sin Y_2}}{1}}
  \right]^2,\\
  \phi_{1+}^{\textrm{2p}}&=-\frac{1}{2}\sin(2\theta_1)\exp(-t\cos\omega_1)
  \frac{\cas{\sin Y_1,\cos Y_1,\eta_1\sin Y_1, \cos Y_2, \eta_2\sin
      Y_2}}
  {\shft{\cas{\cos Y_1, \eta_1\sin Y_1, \cos Y_2, \eta_2 \sin Y_2}}{1}},\\
  \phi_{1-}^{\textrm{2p}}&=-\dot{\beta_1}\exp(t\cos\omega_i)
  \frac{\shft{\cas{\cos Y_1,\cos Y_2, \eta_2\sin Y_2}}{1}} {\cas{\cos
      Y_1, \eta_1\sin Y_1, \cos Y_2, \eta_2\sin Y_2}},
\end{align*}
and $\phi_{2\pm}^{\textrm{2p}}$ have the similar formulae.
$x^{\textrm{2p}}$ describes a wave profile oscillates, decreases to
zero slowly as $|n|\rightarrow\infty$. 
$\phi_{1\pm}^{\textrm{2p}}$ and $\phi_{2\pm}^{\textrm{2p}}$ behave
like $\mathcal{O}(n^{-1})$ as $|n|\rightarrow\infty$. 

The Positon-positon interaction can be analyzed as follows.
Fixing $\eta_1$, assume $|\eta_2|\rightarrow\infty$ when
$t\rightarrow\pm\infty$, then
\begin{align*}
  x^{\textrm{2p}}&\sim \frac{1}{2}\log\left[
    \frac{A(\eta_1+\frac{3}{2})+B\sin(2Y_1+3\omega_1)}
    {A(\eta_1+\frac{5}{2})+B\sin(2Y_1+5\omega_1)} \right]^2,
  \textrm{as }\,t\rightarrow\pm\infty,
\end{align*}
where
\begin{align*}
  A&:=2\sin\omega_1\sin\omega_2+\sin(3\omega_1)\sin\omega_2
  -2\sin(2\omega_1)\sin(2\omega_2)+\sin\omega_1\sin(3\omega_2),\\
  B&:=\sin\omega_2\cos(2\omega_1)+\frac{3}{2}\sin\omega_2
  -\sin(2\omega_2)\cos\omega_1+\frac{1}{2}\sin(3\omega_2).
\end{align*}
Thus, we observe one-positon in terms of $\eta_1$ and $Y_1$ at
$t=\pm\infty$. 
And there is no phase shift during the interaction. Analogous, fixing
$\eta_2$ we observe one-positon at $t=\pm\infty$ in terms of $\eta_2$
and $Y_2$. Similarly, there is no phase shift in the course of
interaction. Positons propagations transparently as if others were
absent. This transparency of interaction is a remarkable feature of
positon solutions. The N-positon solution for TLSCS
\eqref{subeqs:TLSCS} is obtained by taking
$f_i:=\cos\theta_i\exp(-t\cos\omega_i)\cos(n\omega_i-t\sin\omega_i)$,
$g_i:=\sin\theta_i\exp(-t\cos\omega_i)\sin(n\omega_i-t\sin\omega_i)$,
$i=1,\ldots,N$ be solutions of \eqref{subeqs:trivialLaxPair} w.r.t
$\lambda_i=2\cos\omega_i$, $\theta_i\neq\frac{n}{2}\pi$ ($n\in
\mathbb{Z}$), $h_i=f_i+g_i$ and use theorem~\ref{thm:deGFDT} with
$l=0$, $I=N$, $\mathbf{m}=(2,2,\cdots,2)\in\mathbb{N}^N$.


We can construct N-negaton solutions to TLSCS~\eqref{subeqs:TLSCS} by
considering eigenfunctions
$f_i:=\epsilon_i^n\exp(n\omega_i-\epsilon_ie^{\omega_i}t+\theta_i)$,
$g_i:=\epsilon_i^n\exp(-n\omega_i-\epsilon_ie^{-\omega_i}t-\theta_i)$
of \eqref{subeqs:trivialLaxPair} with
$\lambda_i=2\epsilon_i\cosh\omega_i$, distinct
$\omega_i\in\mathbb{R}$, $\theta_i\in\mathbb{R}$, $\epsilon_i=\pm 1$,$h_i=f_i+g_i$ and using theorem~\ref{thm:deGFDT} with
$l=0$, $I=N$, $\mathbf{m}=(2,2,\cdots,2)\in\mathbb{N}^N$.

The GFDT is quite general to construct not only N-soliton
(positon,negaton) solutions, but also solutions of combined type.

\subsection{Positon-negaton solution and interaction}
\label{sec:one-positon-one-negaton}

Let
\begin{align*}
  f_1&:=\cos\theta_1\exp(-t\cos\omega_1)\cos(n\omega_1-t\sin\omega_1),\\
  g_1&:=\sin\theta_1\exp(-t\cos\omega_1)\sin(n\omega_1-t\sin\omega_1),\\
  f_2&:=\epsilon_2^n\exp(n\omega_2-\epsilon_2e^{\omega_2}t+\theta_2),\\
  g_2&:=\epsilon_2^n\exp(-n\omega_2-\epsilon_2e^{-\omega_2}t-\theta_2),\\
\end{align*}
where $\theta_1$, $\theta_2$, $\omega_1$ and $\omega_2$ are real
numbers, $\theta_1\neq k\pi/2 $ for any integer $k$. Define
\begin{displaymath}
  h_1=f_1+g_1=\exp(-t\cos\omega_1)\cos Y_1,\,\,
  h_2=f_2+g_2=2\epsilon_2^n\exp(-t\epsilon_2\cosh\omega_2)\cosh Y_2,
\end{displaymath}
where
\begin{displaymath}
  Y_1:= n\omega_1-t\sin\omega_1-\theta_1,\quad
  Y_2:= n\omega_2-t\epsilon_2\sinh\omega_2 +\theta_2.
\end{displaymath}
Denote
\begin{displaymath}
  \eta_1:=\partial_{\omega_1}Y_1+\frac{1}{2}\beta_1(t)\sin 2\theta_1,\quad
  \eta_2:=\partial_{\omega_2}Y_2+\frac{1}{2}\beta_2(t).
\end{displaymath}
Then according to theorem \ref{thm:deGFDT} with the specification
$l=0$, $I=2$, $\mathbf{m}=(2,2)$, we obtain the positon-negaton
solution for TLSCS \eqref{subeqs:TLSCS}.

To analyze the interaction, we always assume $\beta_1(t)$ and
$\beta_2(t)$ tending to $\mp\infty$ as $n\to\pm\infty$. And all
parameters, including $\omega_1$ $\omega_2$ and $\epsilon_2$ are
positive. 
Fixing $\eta_1$, if for particular $\beta_2(t)$, $\eta_2$ increases
faster than $\sinh(2Y_2)$ (i.e.
$|\eta_2/\sinh(2Y_2)|\rightarrow\infty$) as $t\rightarrow\infty$,
which indicates that negaton travels at a very high speed dominated by
$\beta_2(t)$ then
\begin{displaymath}
  x^{\textrm{pm}}\sim\frac{1}{2}\log\left[
    \frac{\frac{1}{2}\sin(2Y_1+3\omega_1)+\sin\omega_1(\eta_1-\Delta_8)}
    {\frac{1}{2}\sin(2Y_1+5\omega_1)+\sin\omega_1(\eta_1+1-\Delta_8)}\right],
  \quad \textrm{as}\, t\rightarrow \pm\infty,
\end{displaymath}
where
\begin{displaymath}
  \Delta_8:=\sin\omega_1\frac{\cos\omega_1-2\epsilon_2\cosh\omega_2-2\cosh\omega_2\cos\omega_1+2}
  {e^{\omega_2}(\epsilon_2e^{-\omega_2}-e^{i\omega_1})(\epsilon_2e^{-\omega_2}-e^{-i\omega_1})}.
\end{displaymath}
Thus we see the one-positon profile (see (\ref{eq:positon_x})) with
neither phase shift nor displacement at the two end. That is to say
the negaton is transparent for positon, which is a phenomenon
\emph{never} observed in the ordinary Toda lattice case.

If $\eta_2$ increases slower than $\sinh(2Y_2)$
(i.e. $|\eta_2/\sinh(2Y_2)|\rightarrow\infty$) as $|t|$ increases, which
implies a slowly traveling negaton in comparison with the previous
case, then
\begin{displaymath}
  x^{\textrm{pn}}\sim\frac{1}{2}\log\left[\frac
    {\sin\omega_1(\eta_1+\Delta_9)+\frac{1}{2}\sin(2Y_1+\omega_1-\Delta_{11})}
    {\sin\omega_1(\eta_1+1+\Delta_9)+\frac{1}{2}\sin(2Y_1+3\omega_1-\Delta_{11})}\right]^2
  -2\omega_2,\quad
  \textrm{as}\;\; t\rightarrow-\infty,
\end{displaymath}
\begin{displaymath}
  x^{\textrm{pn}}\sim\frac{1}{2}\log\left[\frac
    {\sin\omega_1(\eta_1+\Delta_{10})+\frac{1}{2}\sin(2Y_1+5\omega_1+\Delta_{11})}
    {\sin\omega_1(\eta_1+1+\Delta_{10})+\frac{1}{2}\sin(2Y_1+7\omega_1+\Delta_{11})}\right]^2
  +2\omega_2,\;\;
  \textrm{as}\;\; t\rightarrow+\infty,
\end{displaymath}
where
\begin{align*}
  \Delta_9&:= \frac{1}{2}e^{-2\omega_2}\frac
  {(\epsilon_2e^{\omega_2}-3e^{i\omega_1})(\epsilon_2e^{\omega_2}-3e^{-i\omega_1})-4}
  {(\epsilon_2e^{-\omega_2}-e^{i\omega_1})(\epsilon_2e^{-\omega_2}-e^{-i\omega_1})},\\
  \Delta_{10}&:=\frac{1}{2}e^{2\omega_2}\frac
  {(\epsilon_2e^{-\omega_2}-3e^{i\omega_1})(\epsilon_2e^{-\omega_2}-3e^{-i\omega_1})-4}
  {(\epsilon_2e^{\omega_2}-e^{i\omega_1})(\epsilon_2e^{\omega_2}-e^{-i\omega_1})},
\end{align*}
and
\begin{displaymath}
  \Delta_{11}:=i\log\left(
    \frac{\epsilon_2e^{\omega_2}-e^{i\omega_1}}{\epsilon_2e^{\omega_2}-e^{-i\omega_1}}\right)^2
\end{displaymath}
are all real constants. Thus positon travels with phase shift
determined by $\Delta_9$, $\Delta_{10}$ and $\Delta_{11}$, with
displacement determined by $4\omega_2$ in the course of
collision. This is general phenomenon caused by existence of negaton.

If we fix a coordinate frame which travelling with negaton profile,
then
\begin{displaymath}
  x^{\textrm{pn}}\sim\frac{1}{2}\log\left[
    \frac{\frac{1}{2}\sinh(2Y_2+3\omega_2)+(\eta_2+3)\sinh\omega_2}
    {\frac{1}{2}\sinh(2Y_2+5\omega_2)+(\eta_2+4)\sinh\omega_2}\right]^2
  \quad \textrm{as} \;\;t\rightarrow\pm\infty.
\end{displaymath}
Thus, the negaton travels insensitive about the existence of positon.

Soliton-positon and soliton-negaton solutions can obtained in the same
way by using theorem \ref{thm:deGFDT} with $l=1$, $I=1$,
$\mathbf{m}=2$, $q:=F+\alpha G$, $h:=f_1+g_1$ and $q:=F+\alpha G$,
$h:=f_2+g_2$ respectively. ($f_i,g_i$ $i=1,2$ are defined in the
beginning of this section and $F,G,\alpha$ are defined in the section
\ref{sec:oneSoliton}.)

\section{Conclusions}
\label{sec:conclusions}
Based on the constrained flows of Toda lattice hierarchy, we
constructed Toda lattice hierarchy with self-consistent sources and
their Lax representation.

We developed a method to construct FDT with arbitrary functions of
time and the GFDT with arbitrary functions of time, which, in contrast
with the well-known Darboux transformation for Toda lattice, provide
non-auto-B\"acklund transformation between two TLSCSs with different
degrees and enable us to obtain various explicit solutions to TLSCS.
Resembling the ordinary Toda lattice case, this system possesses
solutions of rich families, including solitons, positons, negatons and
the solutions of combined types. A number of solutions are listed by
our method. The investigation on these solutions shows a quite similar
nature with solutions of ordinary Toda lattice. However, the new
feature concerning interactions between negaton and positon (or
soliton) which is different from the ordinary Toda lattice case is
also stated. This difference is caused by the wide range of variation
of the speed of negatons in TLSCS cases. We note that variation of
speed is a common feature for continuous and discrete systems with
self-consistent sources, see \cite{art:ZengMaLin00,art:LinZengMa01}
etc.

It is convinced that our approach for constructing systems with
self-consistent sources and generalized forward Darboux transformation
technique are available for other discrete systems. Some investigation
will be present in the forthcoming paper.

\section*{Acknowledgments}
  \label{ack}
  This work was supported by the Chinese Basic Research Project
  'Nonlinear Science'.

\end{document}